\documentclass[a4paper,11pt]{article}
\usepackage{pos}

\title{Tunka-Rex Virtual Observatory}
\ShortTitle{ Tunka-Rex Virtual Observatory}

\author*[a]{V. Lenok}
\author[b]{D.~Kostunin}
\author[]{O.~Kopylova}
\author[c]{P.~Bezyazeekov}
\author[a]{D. Wochele}
\author[a]{F. Polgart}
\author[d]{S.~Golovachev}
\author[d]{V.~Sotnikov}
\author[e]{E.~Sotnikova}

\affiliation[a]{Karlsruhe Institute of Technology, Institute for Astroparticle Physics, D-76021 Karlsruhe, Germany}
\affiliation[b]{DESY, 15738 Zeuthen, Germany}
\affiliation[c]{Applied Physics Institute, Irkutsk State University, 664020 Irkutsk, Russia}
\affiliation[d]{JetBrains Research, 194100 St. Petersburg, Russia}
\affiliation[e]{Sobolev Institute of Mathematics, 630090 Novosibirsk, Russia}

\usepackage{todonotes}

\forColl{Tunka-Rex} 

\emailAdd{contact@tunkarex.info}

\abstract{

Tunka-Rex (Tunka Radio Extension) was a detector for ultra-high energy cosmic rays measuring radio emission for air showers in the frequency band of 30-80 MHz, operating in 2010s. It provided an experimental proof that sparse radio arrays can be a cost-effective technique to measure the depth of shower maximum with resolutions competitive to optical detectors. After the decommissioning of Tunka-Rex, as last phase of its lifecycle and following the FAIR (Findability --- Accessibility --- Interoperability --- Reuse) principles, we publish the data and software under free licenses in the frame of the TRVO (Tunka-Rex Virtual Observatory), which is hosted at KIT under the partnership with the KCDC and GRADLCI projects. We present the main features of TRVO, its interface and give an overview of projects, which benefit from its open software and data.
}

\FullConference{37$^{\rm{th}}$ International Cosmic Ray Conference (ICRC 2021)\\
		July 12th -- 23rd, 2021\\
		Online -- Berlin, Germany}

\begin{document}
\maketitle

\section{Introduction}
In the era of large astrophysical facilities, which have a lifecycle exceeding decades and produce a huge amount of heterogeneous data the one of the most important question is related to the conservation of scientific knowledge acquired by them.
The attempts of solution of these issues, which actually relate to the different fields of human knowledge, have resulted into formulation of FAIR (Findability --- Accessibility --- Interoperability --- Reuse) data principles\footnote{\url{https://www.go-fair.org/}}.

One of the most successful example of publication of cosmic-ray data in open access is the KCDC (KASCADE Cosmic Ray Data Centre)\footnote{\url{https://kcdc.iap.kit.edu/}} project~\cite{Haungs:2018xpw}, which contains data from the particle detectors, namely energy deposits and arrival times at each KASCADE-Grande station.
Contrary to particle detectors, digital radio technique is relatively young and has its own specifics~\cite{Schroder:2016hrv} and thus, publication of data from cosmic-ray radio detectors aims at slightly different features.
Since the methods for the radio detectors is actively developed over last years, it is important to provide access to the lower levels of data, i.e. amplitudes of electromagnetic fields on the antennas (one can find the same approach in data releases from gravitational wave\footnote{\url{https://www.gw-openscience.org/}} or fast radio burst detectors\footnote{\url{https://chime-frb-open-data.github.io/}}.

We develop the first framework for the quick and easy access to the data of radio detector designed for the cosmic-ray measurements taking into account features of this particular technique and focusing on the most important requirements from the field.
Tunka-Rex Virtual Observatory (\texttt{TRVO}) is based on this framework and includes radio data from the Tunka-Rex detector\footnote{\url{https://www.iap.kit.edu/tunka-rex/}}~\cite{Bezyazeekov:2015rpa} operating in 2012-2019 in the frame of TAIGA facility~\cite{Kostunin:2019nzy}.
In this work we describe the actual status of \texttt{TRVO}, its development since first prototype~\cite{Bezyazeekov:2019onw}, the structure of the data and architecture of the software.

\section{Tunka-Rex detector and data}
\begin{figure}[t!]
\centering
\includegraphics[width=0.94\linewidth]{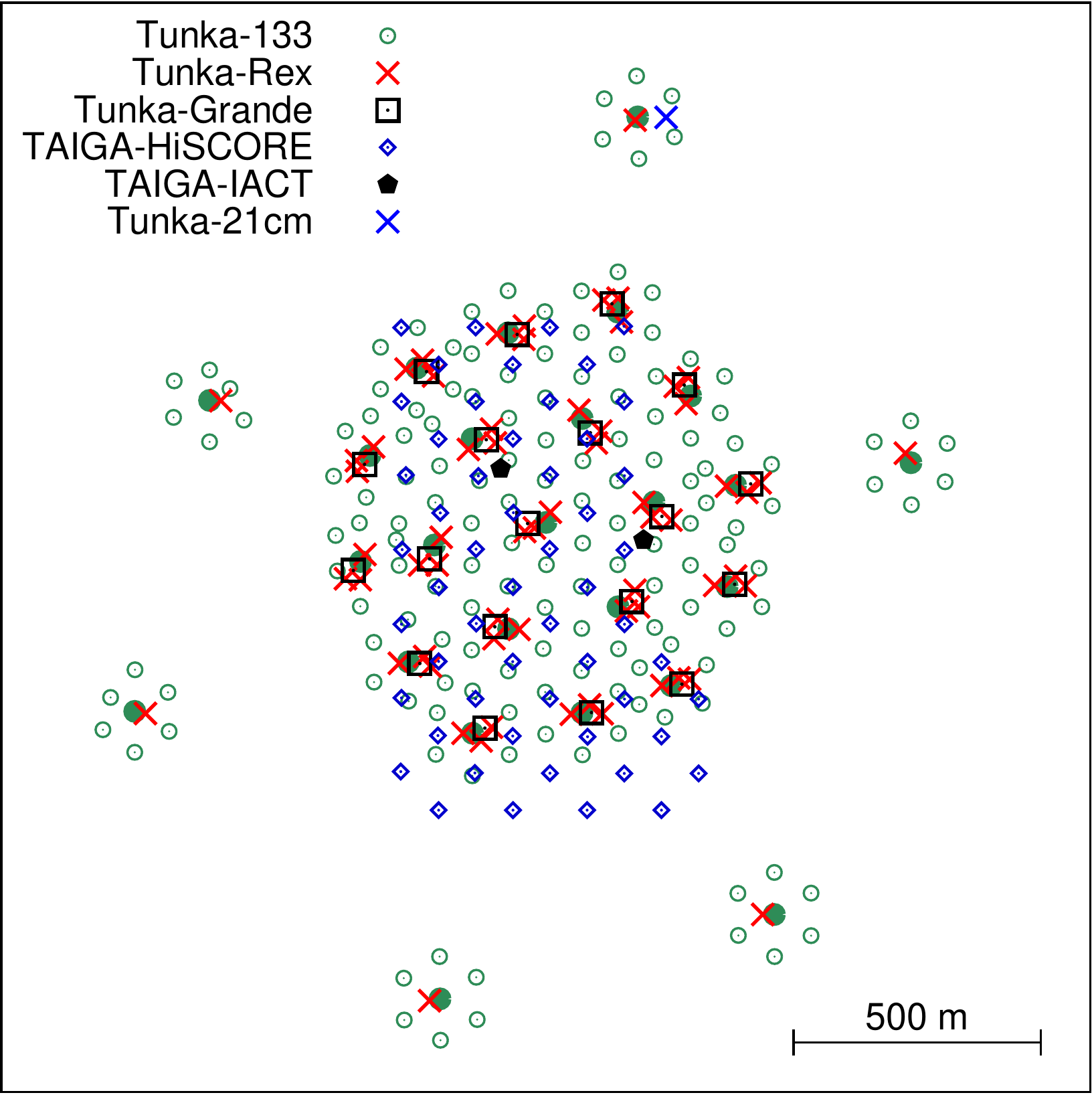}\\
\includegraphics[width=0.94\linewidth]{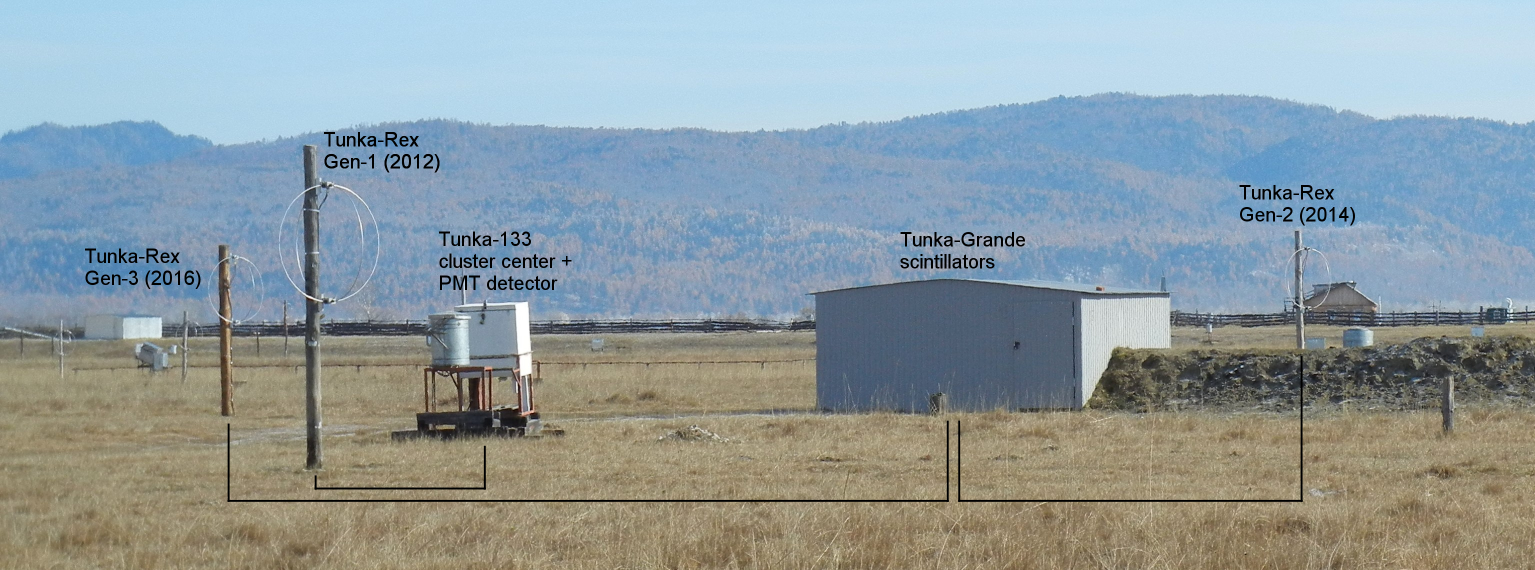}
\caption{
The layout of the TAIGA setups as of 2019.
The antenna stations are depicted as crosses (the Tunka-21cm array is depicted by a single marker due to scale of the map).
The layout and hardware configuration was changes several times during the past years.
\textit{Bottom:} Photo of a single cosmic-ray cluster of the TAIGA facility. 
Lines mark the cable connections.
}
\label{fig:tunka}
\end{figure}

Tunka Radio Extension (Tunka-Rex) was a digital antenna array operating at the Tunka Advanced Instrument for cosmic rays and Gamma Astronomy (TAIGA) observatory in 2012-2019.
In Fig.~\ref{fig:tunka} one can see the layout of the facility before Tunka-Rex decommission and note that the cosmic-ray setups are grouped in clusters: 19 clusters in a dense core and 6 satellite clusters.
Each core cluster is equipped with 3 Tunka-Rex antenna stations, while satellite clusters contain one antenna station, each, and no Tunka-Grande scintillators.

\begin{figure}[t!]
\includegraphics[width=1.0\linewidth]{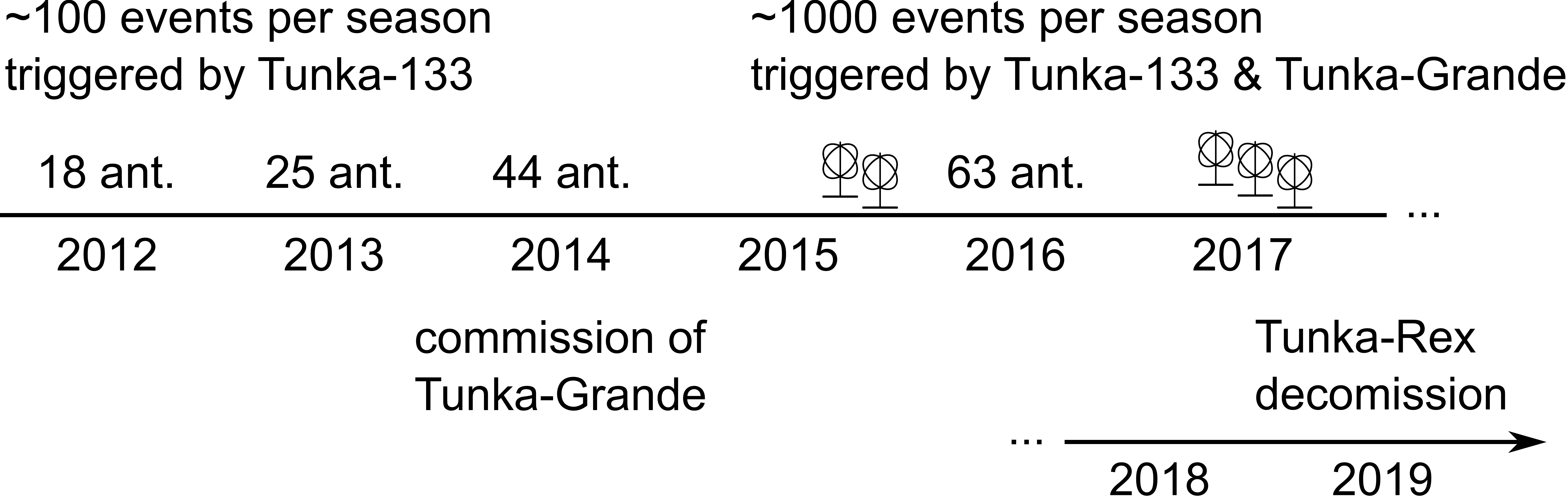}
\caption{The antenna array has been commissioned in 2012 with 18 antenna stations triggered by the Tunka-133 air-Cherenkov detectors.
Since the commissioning of Tunka-Grande in 2014-2015, Tunka-Rex additionally receives a trigger from Tunka-Grande (during daytime measurements).
Starting from 2018, we are working on the public access of Tunka-Rex software and data.}
\label{fig:timeline}
\end{figure}

At its last stage, Tunka-Rex consisted of 57 antenna stations located in the dense core of TAIGA (1~km\textsuperscript{2}) and 6 satellite antenna stations expanding the sensitive area of the array to 3~km\textsuperscript{2}.
Tunka-Rex has been commissioned in 2012 with 18 antenna stations triggered by the air-Cherenkov array Tunka-133.
In the following years, Tunka-Rex was upgraded several times. 
The TAIGA facility was enhanced by the Tunka-Grande scintillator array providing a trigger for Tunka-Rex since 2015.
One can see the timeline of the Tunka-Rex development in Fig.~\ref{fig:timeline}.

\begin{figure}[t!]
\includegraphics[width=1.0\linewidth]{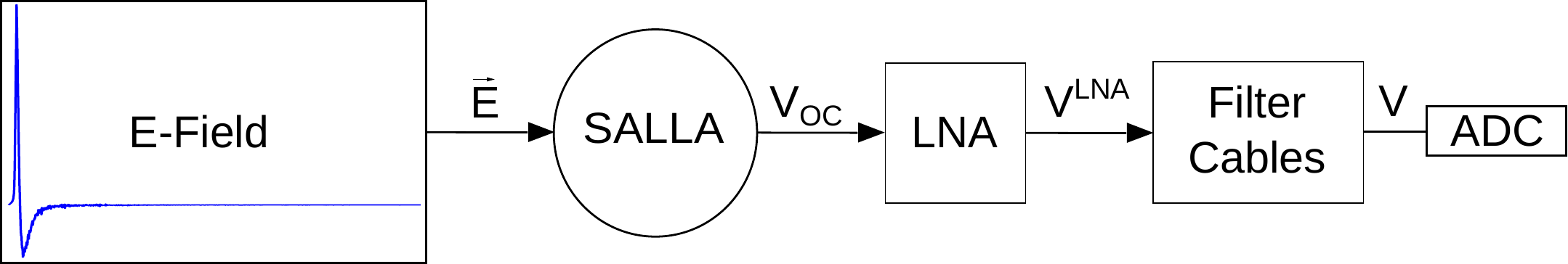}
\caption{Scheme depicting the instrument response function (IRF).
The incoming radio signal is received by the antenna, passes through electronics and cables, and is recorded by the ADC.
For details see Ref.~\cite{Bezyazeekov:2015rpa}.
}
\label{fig:irf}
\end{figure}

Each Tunka-Rex antenna station consists of two perpendicular active Short Aperiodic Loaded Loop Antennas (SALLA)~\cite{Abreu:2012pi} pre-amplified with a Low Noise Amplifier (LNA).
Signals from the antenna arcs are transmitted via 30~m coaxial cables to an analog filter-amplifier, which cuts the frequency band to 30-80~MHz.
The filtered signal is then digitized by the local data acquisition system (DAQ) with a 12 bit-sampling at a rate of 200 MHz; the data are collected in traces of 1024 samples each.
Each element of this signal chain has been calibrated under laboratory conditions, which resulted in the instrument response function (IRF) defining the resulting digital traces recorded by the DAQ (see Fig.~\ref{fig:irf})
For the reconstruction of the original signal, the inverse IRF is convoluted with the raw data.
This convolution defines the \textit{data layers (DL)} explained below.

The distinguishing feature of the broadband radio detectors is that they can be used both for radio astronomy and astroparticle purposes (e.g. ultra-high energy neutrino and cosmic-ray detection) depending on the configuration and operation mode.
For example, the core of the LOFAR antenna array has been successfully applied for cosmic-ray detection~\cite{Schellart:2013bba}; meanwhile the proposed air-shower array GRAND aims also at astronomy goals~\cite{Alvarez-Muniz:2018bhp}.

\section{Implementation}
As for the prototype, the production version of \texttt{TRVO}\footnote{\url{https://gitlab.iap.kit.edu/tunkarex/trvo}} is deployed at IAP KIT.
We have selected PostgreSQL\footnote{\url{https://www.postgresql.org}} for backend storage and implemented communication with database via SQLAlchemy\footnote{\url{https://www.sqlalchemy.org/}}.
All events are stored in the single table, with one row per radio trace (i.e. both channels from single antenna are stored in two rows).
We have developed interface and basic functions for the data selection and analysis, e.g. in the tutorial notebook at the JupyterLab of IAP KIT\footnote{\texttt{tutorials/jbr/trvo/trvo\_icrc.ipynb} at \url{https://jupyter.iap.kit.edu}} on can find the application of autoencoder for the radio traces denoising~\cite{autoencoder}.

\subsection{Structure of the data}
\label{sec:data_structure}
In this section we provide a general description of the Tunka-Rex data types, their structure, and their connection with the hardware of the experiment and observed phenomena.
As described above, raw Tunka-Rex data consist of traces recorded for each antenna from the DAQ buffer after receiving an external trigger.
Below the description of the main table with radio traces:
\begin{itemize}
\item \texttt{uuid}: unique identifier of the entry. We decided to use UUID~\cite{rfc4122} to have an opportunity for combination of data from different detectors without ID overlapping.
\item \texttt{trigger}: trigger configuration of the array, in case of Tunka-Rex it is either Tunka-133 or Tunka-Grande.
\item \texttt{run\_id}: identifier of run, in case of Tunka-Rex we use date in format \texttt{YYYYMMDD}, i.e. unique run is identified by combination of \texttt{trigger} and {run\_id}.
\item \texttt{event\_id}: identifier of the event within single run starting from zero (non-unique identifier)
\item \texttt{station}: identifier of the antenna station, enumerated with the following convention: 1-25 (1st generation), 31-49 (2nd generation), 61-79 (3rd generation)
\item \texttt{timestamp}: float number of the GPS time of the event with nanosecond precision
\item \texttt{traces}: serialized arrays (two channels or three electric-field components) each with at least 1024 elements (non-upsampled Tunka-Rex traces)
\end{itemize}
As will be described below, \texttt{DL0-2} differs only in the way of the representation of the \texttt{traces} field.

\subsection{Calibration and supplementary data}
Antenna\footnote{\url{https://gitlab.iap.kit.edu/tunkarex/antenna}} and electronics\footnote{\url{https://gitlab.iap.kit.edu/tunkarex/calibration}} calibration data and software are published in Tunka-Rex software repository.
We plan integrate this software with \texttt{TRVO} in the nearest future.
The supplementary data (e.g. observation conditions provided by TAIGA meteo) can be extracted via GRADLCI services~\cite{gradlciandy}.

\subsection{Data layers}
In this section we describe the naming conventions for the data layers in the \texttt{TRVO}
\texttt{DL0-2} are organized in the standard structure described above: Station $+$ Calibration $+$ Supplementary data,
while the \texttt{DL3+} can have additional entries, e.g. cosmic-ray events, radio bursts, etc.

\textbf{Data Layer 0} consists of raw traces recorded by the ADCs, i.e. arrays containing values in the range [0;4095].
These data are intended to be used in case of recalibration/debugging of the instrument and are not recommended for the external application.

\textbf{Data Layer 1} consists of the traces containing voltages at the antenna stations (i.e. antenna-induced voltages) obtained after unfolding the raw traces from the hardware response of Tunka-Rex amplifiers, filters, and cables.
From these values the electrical field at the antenna station can be reconstructed using the specific antenna pattern and direction of incoming radio wave.

\textbf{Data Layer 2} consists of the traces containing voltages converted to the values of electrical field at the antenna stations.
Depending on the data release, the electrical fields will be calculated for air-shower events (\texttt{DL2-AIRSHOWER}), for astronomical objects (\texttt{DL2-ASTRONOMY}), or for any other kind of measurements, e.g. background, RFI, etc (\texttt{DL2-OTHER}).

\textbf{Data Layer 3+} contains high-level reconstruction of radio data, i.e. quantities obtained after sophisticated processing and analyzing of radio traces.
These data can be represented in tables, histograms, FITS files, etc.

In the current version of \texttt{TRVO} we provide only \texttt{DL1}, since it does not require recalibration of the raw data on the one hand, and allows for flexible application of antenna pattern on the other one.

\section{Application of the Tunka-Rex Virtual Observatory}
Since the primary goal of Tunka-Rex is the detection of cosmic rays,
the access to the high-level reconstruction of air showers (\texttt{DL3+}) has already been developed in the frame of KCDC/GRADLCI and we do not depart from this concept significantly.
It is worth nothing that \texttt{DL3} radio data can be used for cross-calibration of different cosmic-ray experiments, as shown in Ref.~\cite{Apel:2016gws}.
Below we discuss unique features of the Tunka-Rex archival data and their application to current and future research (one should remember, that the Tunka-Rex trigger is tuned for cosmic-ray detection and the selection from the archival data might be significantly biased and can be used only for tentative studies and prototyping).

\textbf{Studies of the radio background in the frequency band of 30-80~MHz.}
Nowadays there are only few radio telescopes operating in this frequency band, moreover these telescopes operate in an interferometric mode. 
They correlate the radio signal using beam-forming and record the resulting correlation, while radio arrays aimed at cosmic-ray detection record full uncorrelated time series.
The broadband measurement of radio background in this frequency band is of special interest to search for a possible cosmological signal from neutral hydrogen.
Since this signal has a signal-to-noise ratio (SNR) of about $10^{-5}$, understanding of systematic uncertainties is crucial for this type of measurements.
The Tunka-Rex child experiment, Tunka-21cm~\cite{Kostunin:2019gho}, was designed to test the possibility of application of cosmic-ray detectors for studies of this cosmological signal, and is a first user of \texttt{DL2-BACKGROUND} and \texttt{DL2-ASTRONOMY}.
\texttt{TRVO} was also used for the development of self-triggering techniques for the future radio arrays~\cite{selftrigger}.

\textbf{Searching for radio transients.}
Obviously archival data can be used for searching for astronomical transients in this frequency band.
The effective exposure of Tunka-Rex provides only a very small probability of detection of any kind of transients.
However, the archival data can be used for the test of detection techniques for future multi-purpose detectors.

\textbf{Training of neural networks for RFI tagging.}
It was shown, that deep learning can improve the signal reconstruction of radio detectors when using an autoencoder architecture~\cite{Shipilov:2018wph,Erdmann:2019nie,abdul},
because neural networks are able to learn features of the background and can be used for either denoising of radio traces or tagging of traces containing special features.
It is worth noting, that the present Tunka-Rex autoencoder is trained on a dataset containing less than 1\% of all Tunka-Rex background traces, what promises significant improvements by using larger training samples extracted from \texttt{TRVO}.

\textbf{Outreach and education.} 
Open data implies outreach and educational activities, and we support this activities. 
TRVO will be used as educational platform in the outreach part of the KCDC and GRADLCI projects, as well as in others, e.g. in the first worksop of Mathematical center in Akademgorodok\footnote{\url{https://english.nsu.ru/mca/media/news/2973965/}}.
Last but not least, the developed framework can be applied to future arrays, e.g. Tien-Shan cosmic-ray setup~\cite{tienshaneas} and GCOS~\cite{gcos}.

\section{Conclusion}
The Tunka-Rex Virtual Observatory provides open access to the data of experiments measuring cosmic rays with radio technique.
We plan to combine both astroparticle- and astronomy-related features in \texttt{TRVO} and provide fast and user-friendly access with the possibility of custom scripting for complex preselection and preprocessing of the data.

\section*{Acknowledgements}
The authors would like to express gratitude to the colleagues from KCDC team.
The development and testing of the \texttt{TRVO} was supported by the state contract with Institute of Thermophysics SB RAS.

\bibliographystyle{JHEP}
\bibliography{references}

\clearpage
\section*{Full Authors List: \Coll\ Collaboration}
\scriptsize
\noindent
P.~Bezyazeekov$^{1}$,
N.~Budnev$^{1}$,
O.~Fedorov$^{1}$,
O.~Gress$^{1}$,
O.~Grishin$^{1}$,
A.~Haungs$^{2}$,
T.~Huege$^{2,3}$,
Y.~Kazarina$^{1}$,
M.~Kleifges$^{4}$,
E.~Korosteleva$^{5}$,
D.~Kostunin$^{6}$,
L.~Kuzmichev$^{5}$,
V.~Lenok$^{2}$,
N.~Lubsandorzhiev$^{5}$,
S.~Malakhov$^{1}$,
T.~Marshalkina$^{1}$,
R.~Monkhoev$^{1}$,
E.~Osipova$^{5}$,
A.~Pakhorukov$^{1}$,
L.~Pankov$^{1}$,
V.~Prosin$^{5}$,
F.~G.~Schr\"oder$^{2,7}$
D.~Shipilov$^{8}$ and
A.~Zagorodnikov$^{1}$
~\\
~\\
\noindent
$^{1}$Applied Physics Institute ISU, Irkutsk, 664020 Russia\\
$^{2}$Karlsruhe Institute of Technology, Institute for Astroparticle Physics, D-76021 Karlsruhe, Germany\\
$^{3}$Astrophysical Institute, Vrije Universiteit Brussel, Pleinlaan 2, 1050 Brussels, Belgium\\
$^{4}$Institut f\"ur Prozessdatenverarbeitung und Elektronik, Karlsruhe Institute of Technology (KIT), Karlsruhe, 76021 Germany\\
$^{5}$Skobeltsyn Institute of Nuclear Physics MSU, Moscow, 119991 Russia\\
$^{6}$DESY, Zeuthen, 15738 Germany\\
$^{7}$Bartol Research Institute, Department of Physics and Astronomy, University of Delaware, Newark, DE, 19716, USA\\
$^{8}$X5 Retail Group, Moscow, 119049 Russia
\end{document}